\documentclass[aps,prb,twocolumn,floatfix,bibliography,superscriptaddress]{revtex4}
\usepackage{graphics,bm,amsmath,amssymb,natbib,url,epsfig,psfrag,color}
\usepackage{graphicx}
\usepackage{subfigure}
\usepackage[normalem]{ulem}
\begin{document}
	\newcommand{\Tr}{\text{Tr}}
	\newcommand{\beq}{\begin{equation}}
		\newcommand{\eeq}{\end{equation}}
	\newcommand{\eran}[1]{{\color{blue}#1}}
	
	\newcommand{\be}{\begin{equation}}
		\newcommand{\ee}{\end{equation}}
	\newcommand{\bea}{\begin{eqnarray}}
		\newcommand{\eea}{\end{eqnarray}}
	\newcommand{\ES}[1]{{\color{red}{\textbf{{}}}~#1}}
	\newcommand{\sarath}[1]{{\color{blue}{\textbf{{}}}~#1}}
	\newcommand{\CB}[1]{{\color{green}{\textbf{{}}}~#1}} 
	\newcommand{\YM}[1]{{\color{red}#1}}
	
	\def\bs#1\es{\begin{split}#1\end{split}}	\def\bal#1\eal{\begin{align}#1\end{align}}
	\newcommand{\nn}{\nonumber}
	\newcommand{\sgn}{\text{sgn}}
	\title{Detector-tuned overlap catastrophe in quantum dots}

	\author{Sarath Sankar}
	\affiliation{School of Physics and Astronomy, Tel Aviv University, Tel Aviv 6997801, Israel}
	\author{Corentin Bertrand}
	\affiliation{Center for Computational Quantum Physics, Flatiron Institute, New York, New York 10010, USA}
	\author{Antoine Georges}
	\affiliation{Coll\`ege de France, PSL University, 11 place Marcelin Berthelot, 75005 Paris, France}	
	\affiliation{Center for Computational Quantum Physics, Flatiron Institute, New York, New York 10010, USA}
	\affiliation{Department of Quantum Matter Physics, University of Geneva, 24 quai Ernest-Ansermet, 1211 Geneva, Switzerland}
	\affiliation{CPHT, CNRS, Ecole Polytechnique, IP Paris, F-91128 Palaiseau, France}
	\author{Eran Sela}
	\affiliation{School of Physics and Astronomy, Tel Aviv University, Tel Aviv 6997801, Israel}
	
	\author{Yigal Meir}
	\affiliation{Department of Physics, Ben-Gurion University of the Negev, Beer-Sheva 84105, Israel}

	\date{\today}
	\begin{abstract}
		The Anderson overlap catastrophe (AOC) is a many-body effect arising as a result of a shakeup of a Fermi sea due to an abrupt change of a local potential, leading to a power-law dependence of the density of states on energy. Here we demonstrate that a standard quantum-dot detector can be employed as a highly tunable probe of the AOC, where the power law can be continuously modified by a gate voltage. We 
		show 
		that signatures of the AOC have already appeared in previous experiments, and give explicit predictions allowing to tune and pinpoint their nonperturbative aspects. 
	\end{abstract}
	\maketitle
	\section{Introduction}
	Quantum dots (QDs) are highly-tunable mesoscopic systems, which allow detailed exploration of many-body effects in electronic systems. The quintessential example is the demonstration of the Kondo effect in transport through a QD \cite{goldhaber1998kondonature,goldhaber1998kondoprl}, where, by tuning the dot energy level, one can explore the localized moment, mixed valence, and the Kondo regimes in a single sample. Another many-body effect that has been explored in such a system is the Anderson overlap catastrophe (AOC)~\cite{anderson1967infrared}, where a sudden change in the local potential seen by a Fermi sea changes the ground state of the system into another one, such that its overlap with the initial ground-state decays as a power-law of the number of electrons. This effect was first discovered in the context of x-ray absorption in metals \cite{mahan1967excitons,nozieres1969singularities}. 
	Manifestation of the AOC as a Fermi edge singularity in the transport through a QD has been pointed out early on \cite{matveev1992interaction}, and has been extensively explored theoretically \cite{bascones2000nonequilibrium,muzykantskii2003fermi,abanin2004tunable,hentschel2005fermi, wunsch2008electron,goldstein2010population,mkhitaryan2011fermi,chernii2014fermi,goremykina2017fermi,borin2017manifestation}.
	Indeed, signatures of Fermi-edge singularity have been reported in tunneling experiments through single levels and QDs \cite{geim1994fermi,benedict1998fermi,thornton1998many,hapke2000magnetic,frahm2006fermi,ruth2008fermi,krahenmann2017fermi} and also in optical absorption of QDs \cite{latta2011quantum}.
	However, none of these studies have demonstrated the desired tunability expected in QD structures. 
	
	Here we propose using a nearby QD detector (QDD)
	to continuously tune the overlap catastrophe exponent. While it was suggested in \cite{abanin2004tunable}  that one may tune this parameter by modifying the QD shape or its scattering matrix, such modifications are experimentally nontrivial. In our approach one merely has to change the QDD level position, easily controllable by an appropriate gate voltage, thus allowing continuous modification of the AOC exponent. Such quantum-point-contact or QD detectors have been employed extensively in the last 30 years~\cite{field1993measurements}, and thus the setups we explore here [see, e.g.,  Fig.~\ref{fig:setup}(a)] are readily realizable. Moreover, as the bias voltage on the QDD can be easily changed, one can also probe experimentally the AOC out of equilibrium. We demonstrate below that the puzzles raised in such a recent experiment \cite{ferguson2023measurement} can be quantitatively explained by our approach, and offer, in fact, initial indication of AOC physics in these systems.  We make specific predictions of unequivocal signatures of the AOC in these devices.
	
	The basic setup  consists of a mesoscopic system, comprising of one or several QDs, where some of them are coupled electrostatically to a QDD, tunnel coupled to its own leads, see Fig.~\ref{fig:setup}(a). When an electron tunnels to a QD that is coupled to the QDD, it shifts the QDD energy, and this modifies the current through it. Accordingly, the QDD serves as a charge detector. 
	It was already pointed out in \cite{aleiner1997dephasing} (see also \cite{goldstein2010population}) that such a system displays AOC physics since the scattering potential of the QDD changes abruptly when an electron tunnels into the mesoscopic system. The corresponding power-law exponent depends on the change in the phase shift of the scattering amplitude through the QDD. Since this phase shift depends continuously on the QDD energy position, the latter can be tuned to change the power-law exponent.
	
	The standard approach to study the effect of the detector on the system - the measurement backaction (MBA) - is the popular $P(E)$ theory~\cite{ingold1992charge,aguado2000double,gustavsson2007frequency}, describing photon assisted tunneling~\cite{kouwenhoven1994observation,platero2004photon}, where the 
	current fluctuations in the detector 
	are considered as an electro-magnetic environment which exchanges energy with the system. 
	However, the $P(E)$ approach fails to take into account the nonperturbative effect of the system state on the environment modes, which is the one leading to AOC. 
	Below, we present an exact treatment of the electrostatic interaction [
	$\lambda$ in Fig.~\ref{fig:setup}(a)] by numerically implementing the approach described in Ref.~\onlinecite{nozieres1969singularities}. We also show that the $P(E)$ approach 
	is a perturbative limit of our theory.
	
	\section{General formalism}
\begin{figure}[t]
		\includegraphics[width=\columnwidth]{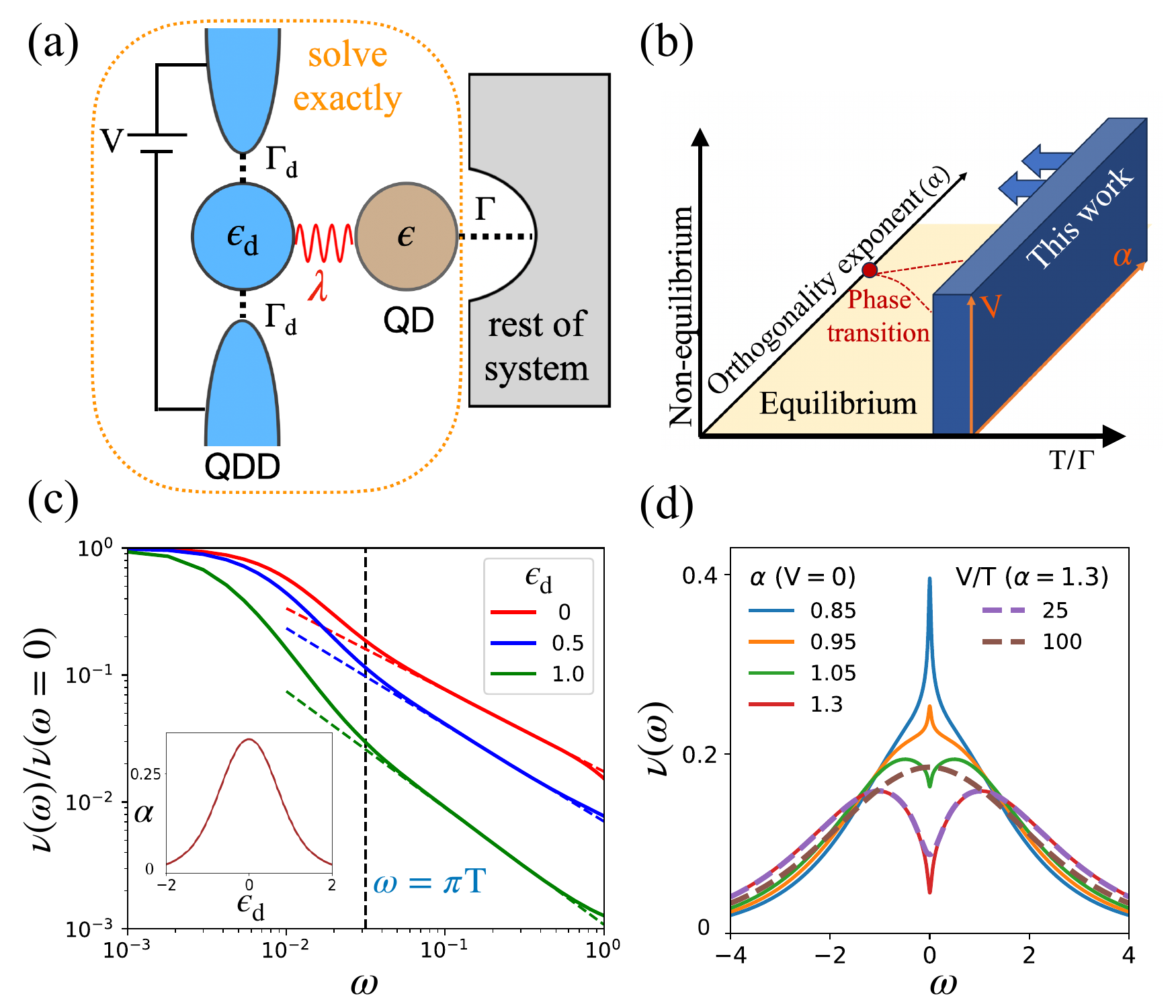}
		\caption{\label{fig:setup} (a) Detector-system setup: The system comprises of a QD  capacitively coupled ($\lambda$) to a detector and tunnel coupled ($\Gamma$) to other QDs or leads. $P(E)$ theory is 
			perturbation 
			in $\lambda$. Here we treat $\lambda$ exactly, together with the bias $V$ within the encircled dashed region, and perform an expansion in $\Gamma/T$. (b) 
			Our theory gives a high-temperature perspective to the role of MBA including nonequilibrium effects at finite bias $V$. (c) Tuning of AOC exponent $\alpha$ in the DoS of the QD via $\epsilon_{\rm{d}}$, $\rho(\omega)\sim |\omega|^{\alpha-1}$  (we set $\Gamma_{\rm{d}}=1$ thoughout the paper) for $ T \ll |\omega| \ll 1$. Here $V=0,\,\lambda=1.5 ,\, T=0.01$, and the degeneracy factor is $m=2$. The inset shows the dependence of $\alpha$ on $\epsilon_d$ as given  by Eq.~(\ref{eq:phase_shift}). (d)  DoS of the QD at $\langle n \rangle=1/2$ and large $\alpha$ for $T=0.01$, $V=0$ and $\epsilon_{d}=0$ (\textit{solid lines}). The DoS splits (two peaks) for $\alpha>1$. Increasing $V$ (\textit{dashed lines}) suppresses this splitting. 
		} 
	\end{figure}

	The main idea 
	is to first, following Refs.~\onlinecite{nozieres1969singularities, ng1996fermi, aleiner1997dephasing}, treat exactly the
	detector, consisting of two voltage-biased leads and a central scattering region, that is capacitively coupled to the system QD [orange contour in  Fig.~\ref{fig:setup}(a)]. Then we treat the tunneling to the rest of the system, $\Gamma$
	, perturbatively. In other words, as displayed in Fig.~\ref{fig:setup}(b), we consider high temperature $T$ compared to quantum coherent scales associated with $\Gamma$ (such as Kondo effects).
	
	Since the system QD has no dynamics at $\Gamma=0$, we can, without loss of generality, describe it by a single spinless level with energy $\epsilon$. The dynamics of the many-body state of the detector is then governed by some  Hamiltonian, $H_n$, that depends on the charge 
	of the QD, $n=\{0,1\}$. The detector affects the 
	rates of electron tunneling in and out of the QD to the rest of the system, $\Gamma^{{{\rm{(in/out)}}}}$, due to inelastic processes. The ``rest of the system" in Fig.~\ref{fig:setup}(a) is arbitrary and various examples are considered below. As a representative case, we now consider it to be a lead whose bare tunneling rate to the QD is $\Gamma$. In the weak tunneling regime $\Gamma \ll T$
	, the 
	tunnelling rates 
	can be written as (see Appendix \ref{sec:app_a} for a derivation based on the Fermi golden rule)
	\bea
	\label{eq:tun_rates}
	\Gamma^{({\rm{in}})}&=&\Gamma \int dE \, f(E+\epsilon)A^{+}(E),~ \nonumber \\ \Gamma^{({\rm{out}})}&=&\Gamma \int dE\, [1-f(E+\epsilon)]A^-(E),
	\eea
	where $f(E)$ is the Fermi function of 
	the lead and we set $\hbar, k_B=1$. The functions $A^{\pm}(E)$ are independent of what the ``rest of the system" is, and  $A^{\pm}(E)=\int dt e^{i E t} A^{\pm}(t)$, with 
	\begin{equation}
		\label{eq:a_corr_time}
		A^{\text{\tiny +/}\text{\small -}}(t)={\rm{Tr}}\left[\rho_{0/1}e^{itH_0}e^{-itH_1}\right],
	\end{equation}
	where $\rho_n \propto e^{- H_n/T} $ denote the detector density matrix for QD charge $n$. The functions $A^\pm (t)$ describe the effect of a sudden change of the Hamiltonian of the detector, and their Fourier transforms $A^\pm (E)$ give the probability of the detector to absorb/emit energy $E$ in the tunneling-in/out process. 
	
	The QDD has a Lorentzian density of states associaed with the $\Gamma_{\rm{d}}-$broadened 
	level. 
	For a detector with constant density of states, explicit expressions were derived in the wide-band limit (WBL)~\cite{aleiner1997dephasing} (also see Appendix \ref{sec:aap_C}); 
	At $T=0$ 
	one has $A_{{\rm{WBL}}}^{\pm}(t) \sim 1/t^\alpha\times e^{-\Gamma_\varphi t}$, 
	where $\alpha$ is the AOC exponent and $\Gamma_\varphi = \gamma V$ is a voltage induced dephasing.
	
	$P(E)$-theory also expresses tunneling rates within the system in terms of a function characterizing the detector~\cite{aguado2000double}. It is important to compare it with our formalism: denoting $A^\pm(t)=e^{C^\pm (t)}$, $P(E)$ theory corresponds to a perturbative approximation of $C^\pm(t)$ to second order in  $\lambda$ (see Appendix \ref{sec:app_D}). 
	It thus includes the MBA 
	of the detector on the system but ignores the back-reaction of the system on the detector, which lies at the heart of the AOC.
	
	Either within 
	$P(E)$ theory or within the WBL,
	the MBA is described by a single function
	\be
	\label{eq:P(E)}
	A^+(E) = A^-(-E) \equiv P(E).
	\ee
	Our treatment of AOC 
	in general, gives $A^+(E) \ne A^-(-E)$. 
	As we show below, this inequality has smoking gun signatures in experiments.
	
	\subsection{ QD properties for $\Gamma \to 0$ \label{sec:app_A_4}}
	All the physically relevant quantities associated with the weakly tunnel coupled QD can be expressed in terms of the $A^{\pm}$ correlators which thus carry all the information regarding MBA.  For example, the QD occupation probability, $\langle n \rangle$, is determined by imposing steady state to the rate equation. For instance, in the QD-lead system, the steady state condition is
	\begin{equation}
		\label{eq:occ_dot_lead}
		(1-\langle n \rangle)\Gamma^{({\rm{in}})}+\langle n \rangle \Gamma^{({\rm{out}})}=0\implies \langle n \rangle=\frac{\Gamma^{({\rm{in}})}}{\Gamma^{({\rm{in}})}+\Gamma^{({\rm{out}})}}.
	\end{equation}
	
	The Green functions 
	of the QD
	can also be expressed in terms of $A^{\pm}(t)$. In particular  the Green functions 
	$G^<(t)=i\langle c^\dagger c(t)\rangle$ and 
	$G^>(t)=-i\langle c(t) c^\dagger \rangle$ 
	are given by
	\begin{equation}
		G^{>/<}(t)=i \left(\langle n \rangle-\frac{1}{2}\mp\frac{1}{2}\right)e^{-i\epsilon t}A^{\pm}(t),
	\end{equation}
	and the retarded Green function can be obtained using $G^R(t)=\theta(t)\left[G^>(t)-G^<(t)\right]$. Note that in the frequency domain, the function $A_R^{\pm}(\omega)$, which is the Fourier transform of $\Theta(t)A^{\pm}(t)$ enters into the expression of $G^R(\omega)$. Since by definition we have, $A^{\pm}(-t)=\left(A^{\pm}(t)\right)^*$,  it follows that $A^{\pm}(\omega)=2{\rm{Re}}\left[A^{\pm}_{R}(\omega)\right]$. One can then express the DoS, $\nu(\omega)=-(1/\pi){\rm{Im}}[G^R(\omega)]$, as
	\begin{equation}
		\label{eq:dos}
		\nu(\omega;\epsilon)= \frac{1}{2\pi}\left[(1-\langle n \rangle)A^+(\omega-\epsilon)+\langle n \rangle A^-(\omega-\epsilon)\right].
	\end{equation}
	This expression for the density of states is used in Fig.~1(c,d).

	
	
	\section{Model} 
	We consider the Hamiltonian, 
	$H=H_{\rm{{sys}}}+H_{\rm{{det}}}+H_\lambda$. The system Hamiltonian $H_{\rm{{sys}}}$  describes a QD 
	with charge $n=0,1$, and the ``rest of the system" %
	in Fig.~\ref{fig:setup}(a),
	and will be specified in the examples below. 
	The detector consists of a QDD with an energy level $\epsilon_{d}$ tunnel coupled to two leads $(\mu=L,R)$ at bias voltage $\pm V/2$ and constant density of states (DoS) $\nu_d$. Its $n-$dependent Hamiltonian is denoted as $H_n=H_{\rm{{det}}}+H_\lambda$,  
	and is given by
	\begin{align}
		\label{eq:model}
		H_n&=\sum_{i=1}^m [\epsilon_{{\rm{d}}} d_{i}^\dagger d_{i} +\sum_{\mu=L,R} (\epsilon_k \varphi_{ki\mu}^\dagger \varphi_{ki\mu} + v_\mu \varphi_{ki\mu}^\dagger d_{i} +h.c. ) ] \nonumber\\&+\lambda\sum_{i=1}^{m}\left(d_{i}^\dagger d_{i}-\frac{1}{2}\right)\left(n-\frac{1}{2}\right),
	\end{align}
	where $d_{i}^\dagger$ creates an electron in the QDD, and $\varphi_{ki\mu}^\dagger$ creates an electron with momentum $k$ in lead $\mu$.
	Here $i$ denotes additional indices, which can include spin and other channels, and for simplicity, we treat it as a degeneracy factor with $m$ flavors. 
	The total tunneling width of the QDD is $\Gamma_{\rm{d}}=2\pi \nu_d (v_L^2+v_R^2)$.
	The model Eq.~(\ref{eq:model}) ignores intra-detector dot interactions to elucidate AOC physics in the simplest way, as discussed below. A treatment of AOC effects in an interacting detector is left for future work.

	Since $H_n$ is noninteracting, the  correlators $A^{\pm}(E)$ can be obtained exactly. Based on the method of Ref.~\onlinecite{nozieres1969singularities}, 
	we obtain an exact numerical solution 
	focusing on the QDD with finite $V$ and $T$ (see Appendix \ref{sec:app_B} for details on the numerical method).
	Particularly, the AOC exponent  is given by $\alpha=m(\delta/\pi)^2$ where
	\begin{equation}
		\label{eq:phase_shift}
		\delta = \arctan\left(\frac{\epsilon_{{\rm{d}}}+\lambda/2 }{\Gamma_{{\rm{d}}}}\right)-\arctan\left(\frac{\epsilon_{{\rm{d}}}-\lambda/2 }{\Gamma_{ {\rm{d}} }}\right),
	\end{equation}
	so it can be continuously modified  by changing $\epsilon_{\rm{d}}$, as seen in the inset of Fig.~\ref{fig:setup}(c). Pushing $\alpha$ to large values  
	requires either $\lambda \gg \Gamma_{{\rm{d}}}$ or 
	multiple transverse degrees of freedom, increasing $m$.  
	
	We emphasize that $A^\pm(E)$ 
	determines the 
	QD properties at $\Gamma/T  \ll 1$, such as its DoS Eq.~(\ref{eq:dos}). 
	As we can see in Fig.~\ref{fig:setup}(c), the DoS at $\epsilon=0$ for $\langle n \rangle=1/2$ is given by $\nu(\omega;0) \sim |\omega|^{\alpha-1}$ for $T \ll \omega \ll \Gamma_{\rm{d}}$, and transitions from a peak to a dip as $\alpha$ exceeds unity [see Fig.~\ref{fig:setup}(d)].  Since the QD DoS can be easily measured \cite{de2002out,leturcq2005probing},  a straightforward verification of the AOC physics in QDs would be the observation of such power-law DoS, which changes with the QDD energy.  Below, we show that signatures of AOC physics have already appeared in previous experiments, and describe additional smoking-gun measurements.
	
	\begin{figure}[t]
		\includegraphics[width=\columnwidth]{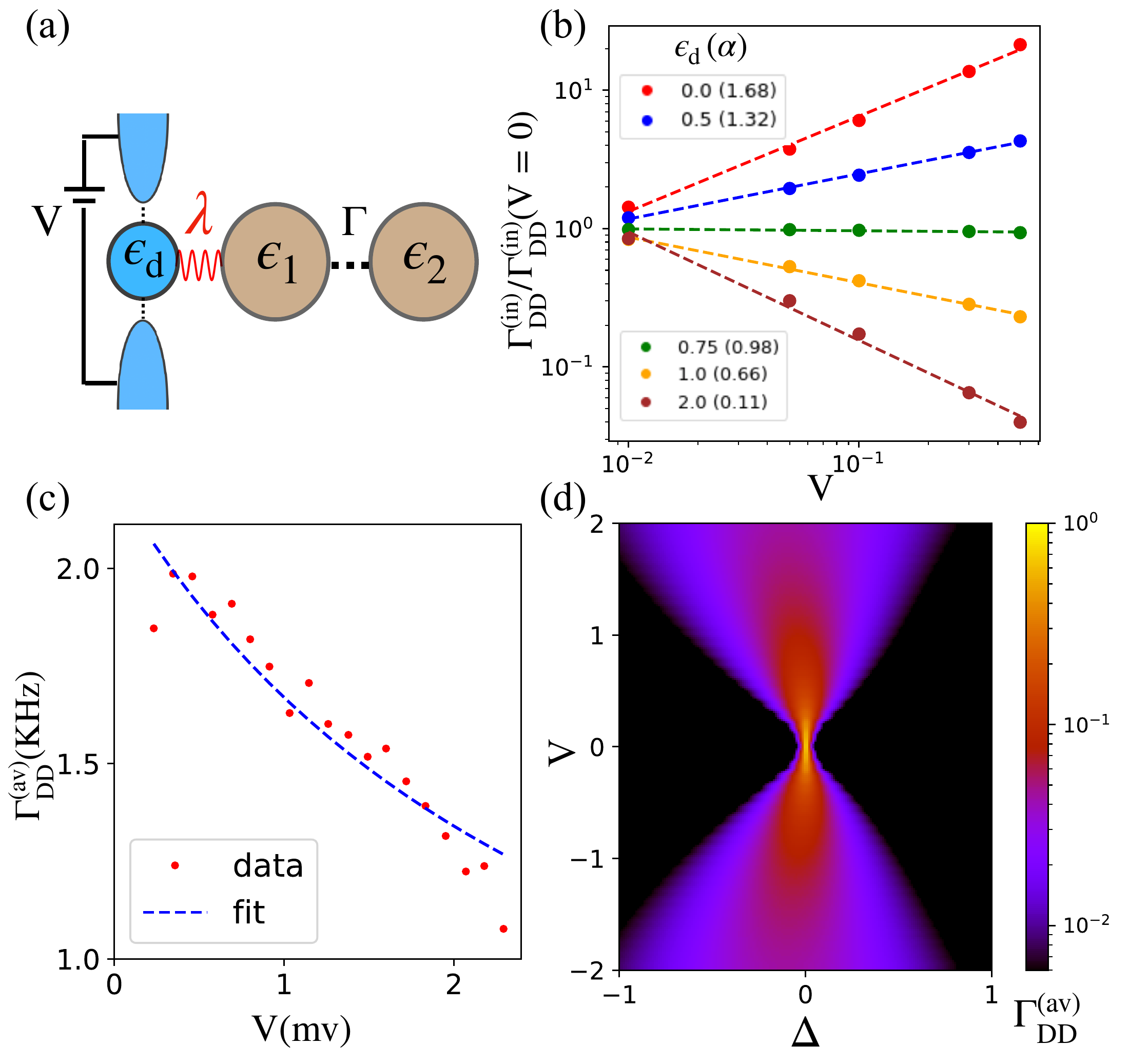}
		\caption{\label{fig:tunratedos}  AOC in a tunnel coupled DD 
			shown in (a) that was studied in Ref.~\onlinecite{kung2009noise}. (b) Dependence of $\Gamma_{\rm{DD}}^{{\rm{(in)}}}$ 
			at zero detuning on $V$ for different $\epsilon_{\rm{d}}$, and thereby different $\alpha$ [see Eq.~\eqref{eq:phase_shift}], $\lambda=1.5, m=10$ and $T=0.002$ obtained by the numerical calculation (\textit{solid points}). 
			For the $T \ll V < \Gamma_{\rm{d}}$ region
			shown here, it follows from Eq.~\eqref{eq:agreat_anlyt_v} and $\Gamma_\varphi \propto V$ that $\Gamma_{\rm{DD}}^{{\rm{(in)}}}(V)\sim V^{\alpha-1}$ (\textit{dashed lines}). (c) Analysis of the zero detuning data in Ref.~\onlinecite{kung2009noise}. The fit is to the analytical form in Eq.~\eqref{eq:agreat_anlyt_v}  with $\alpha=0.05$ and $\Gamma_\varphi=\Gamma^0_\varphi+\gamma V$ with $\Gamma^0_\varphi=0.15$ and $\gamma=0.06$.  (d) Average event rate versus $\Delta$ and $V$ [
			same parameters as in (b)]. Asymmetry with respect to $\Delta \to -\Delta$ captures AOC beyond $P(E)$ theory. }
	\end{figure}
	\section{Tunnel-coupled double dot (DD)} 
	Consider the experimental setup of Kung \emph{et. al.}~\cite{kung2009noise} 
	as shown in  Fig.~\ref{fig:tunratedos}(a),  described by 
	\be
	H_{{\rm{sys}}}=
	\epsilon_1 c^\dagger_1 c_1+\epsilon_2 c^\dagger_2 c_2+ (v c^\dagger_1 c_2 + h.c.),
	\ee
	with $n=c^\dagger_1 c_1$. The  in/out tunneling rates  of QD1 $ \Gamma_{\rm{DD}}^{({\rm{in}}/{\rm{out}})}$ have been measured through real-time charge detection~\cite{kung2009noise}.
	All observations can be accounted for by our theory, using the relations (see Appendix \ref{sec:app_a} for a derivation based on the Fermi golden rule) 
	\be
	\label{eq:rates_dd}
	\Gamma^{({\rm{in}})}_{{\rm{DD}}}=|v|^2 A^+(\epsilon_2-\epsilon_1),~~\Gamma^{({\rm{out}})}_{{\rm{DD}}}=|v|^2 A^-(\epsilon_2-\epsilon_1).
	\ee
	As shown in Fig.~\ref{fig:tunratedos}(b),   at zero detuning, $\Delta \equiv \epsilon_2-\epsilon_1=0$,  the tunneling rates
	have a power-law dependence on  $V$, $\Gamma^{({\rm{in}})}_{{\rm{DD}}} \propto V^{\alpha-1}$ for $V\gg T$, reflecting the power-law dependence of the DoS. 
	Thus, this measurement can be used to gauge the orthogonality exponent $\alpha$, see Fig.~\ref{fig:tunratedos}(b).

	Following Kung \emph{et. al.}~\cite{kung2009noise} we define the event rate - the harmonic average of the tunneling rates, $1/\Gamma_{\rm{DD}}^{\rm{(av)}}(\Delta) = 1/\Gamma_{\rm{DD}}^{{(\rm{in})}}(\Delta)+1/\Gamma_{\rm{DD}}^{{(\rm{out})}}(\Delta)$, 
	determined by the slower rate.
	In Fig.~\ref{fig:tunratedos}(c) we depict the good agreement between  the experimental data for the voltage dependence of $\Gamma_{\rm{DD}}^{\rm{(av)}}(\Delta=0)$
	and the results of our model. We fit the data using the WBL 
	of $A^\pm(\omega)$ derived in Appendix \ref{sec:aap_C}:
	\begin{equation}
		\label{eq:agreat_anlyt_v}
		A^{\pm}_{{\rm{WBL}}}(\omega) 
		\cong \chi(\alpha)
		{\rm{Re}}\left[\frac{e^{\pm i\pi\alpha/2}\Gamma(\frac{i\omega+\Gamma_\varphi}{2\pi T}+\frac{\alpha}{2})}{\Gamma(1+\frac{i\omega+\Gamma_\varphi}{2\pi T}-\frac{\alpha}{2})}\right],
	\end{equation}
	with $\chi(\alpha)=\frac{2^\alpha}{\xi}\left(\frac{\pi T}{\xi}\right)^{\alpha-1}\Gamma(1-\alpha)$ and $\xi \sim \Gamma_{\rm{d}}$ a high energy cutoff.
	To obtain the fit we assume that $\Gamma_\varphi=\Gamma^0_\varphi+\gamma V$, where $\Gamma_\varphi^0$ phenomenologically accounts for the energy relaxation by phonons that is present in the experiment. We find a small value for $\alpha \simeq 0.05$, probably due to the weak interaction between the QD and the detector. We predict that beyond a critical $\alpha$, $\alpha=1$, $\Gamma_{\rm{DD}}^{\rm{(av)}}(\Delta=0)$ becomes an increasing function of $V$.
	
	From Eq.~(\ref{eq:rates_dd}), it follows that the event rate is determined by $A^+(\Delta)$ when $\Delta<0$, and by $A^-(-\Delta)$ when $\Delta>0$. 
	The difference between $A^+(\Delta)$ and $A^-(-\Delta)$, a signature of AOC physics, shows up as an asymmetry around $\Delta=0$ in the event rate color plot, see Fig.~\ref{fig:tunratedos}(d). 
	
	\section{Measurement induced population switching (MIPS)} 
\begin{figure}[t]
		\includegraphics[width=\columnwidth]{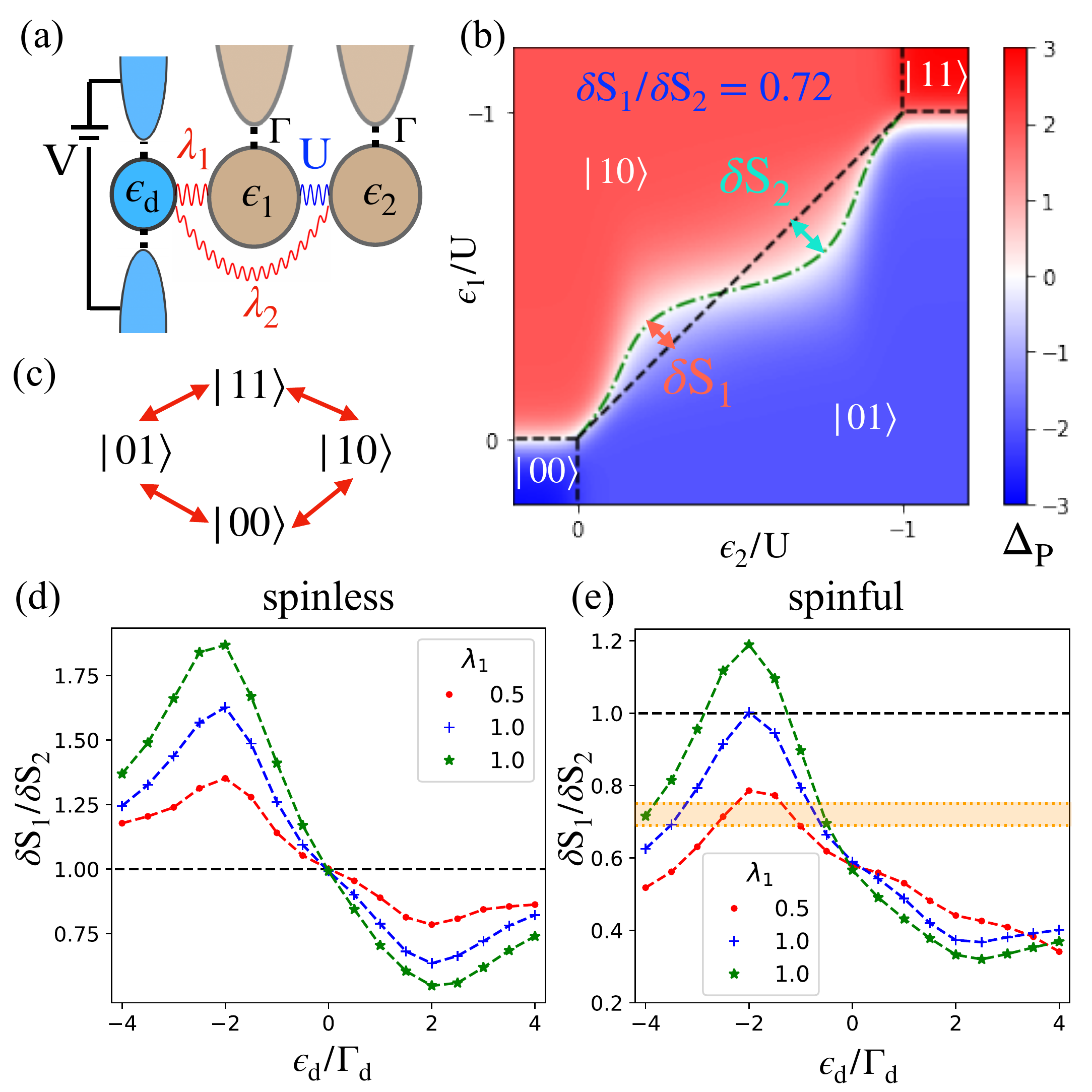}
		\caption{\label{fig:mips}
			(a) 
			DD with no interdot tunneling studied in Ref.~\onlinecite{ferguson2023measurement}.  (b) Charge stability diagram for $\lambda_{1,2}=0$ (\textit{black dashed lines}) and for $\lambda_{1,2} \ne 0$ where we  plot $\Delta_{\rm{P}}=3P_{11}+2P_{10}-2P_{01}-3P_{00}$ and the contour $P_{10}=P_{01}$ (\textit{green dash-dot curve}). 
			Here $U=V=1,\, \epsilon_{d}=0.25,\,\lambda_1=2\lambda_2=0.5,\,T=0.025$, 
			chosen to be close to those reported in Ref.~\onlinecite{ferguson2023measurement} (see Appendix \ref{sec:app_E}). The asymmetry of the S-shape is $\delta S_1/\delta S_2=0.72$, very close to the experimentally reported one [shaded region in (e)]. 
			(c) Transitions between charge states $(n_1 n_2)$ are facilitated by tunneling $(\Gamma)$ into leads.
			(d,e) Predictions for the asymmetry in the spinless and spinful cases. 
			Marked points are numerical results and the dashed lines are guides to the eye. Any asymmetry $\delta S_1/\delta S_2 \ne 1$ in the spinless case and any flipped asymmetry $\delta S_1/\delta S_2>1$ in the spinful case, are 
			signatures of AOC beyond $P(E)$ theory, (i.e. $A^+(E) \ne A^-(-E)$). Here $\lambda_2=0$, $U=1,\,V=3,\,T=0.02$.
		}
	\end{figure}

We now demonstrate that the above microscopic theory explains quantitatively the results and the puzzle raised in a recent elegant experiment~\cite{ferguson2023measurement} studying 
	a capacitively coupled DD. 
	Each dot 
	is tunnel coupled to its respective lead, with dot 1 more strongly coupled to a charge detector than dot 2, see Fig.~\ref{fig:mips}(a). The Hamiltonian 
	is 
	\be
	H_{\rm{sys}}= H_{{\rm{DD}}} + H_{{\rm{leads}}}+ H_{{\rm{tun}}},
	\ee
	with 
	$ H_{{\rm{DD}}}=\sum_\sigma(\epsilon_1 c^\dagger_{1\sigma}c_{1\sigma}+\epsilon_2 c^\dagger_{2\sigma}c_{2\sigma})+U \hat{n}_1 \hat{n}_2+ W (\hat{n}_1^2+\hat{n}_2^2)$, where $c^\dagger_{i\sigma}$ creates an electron with spin $\sigma$ on dot $i$, and $\hat{n}_i=\sum_{\sigma}c^\dagger_{i\sigma}c_{i\sigma}$. Unlike the experiment discussed in the previous paragraph, here there is no 
	tunneling between the dots. 
	The capacitive interaction between the QDD and the DD is described by 
	$H_{{\rm{int}}}=\sum_{i=1}^m\left(d^\dagger_{i}d_{i}-1/2\right) \sum_{\ell=1,2} \lambda_\ell \left( \hat{n}_\ell-1/2\right)$.  The leads and tunneling Hamiltonians are $H_{{\rm{leads}}}= \sum_{\ell k\sigma} \epsilon_k \psi_{\ell k\sigma}^\dagger \psi_{ \ell k\sigma}$,  $H_{{\rm{tun}}}=\sum_{\ell k\sigma} \left(v \psi_{\ell k\sigma}^\dagger c_{\ell  \sigma} +h.c.\right)$, 
	with $\Gamma=2\pi \nu v^2$.
	
	The experiment is described by 
	a large Coulomb interaction $W$ within each QD such that double occupancy is prohibited. Then there are 
	four 
	charge states denoted $\{n_1 n_2\}=\{00,10,01,11 \}$. For $\lambda_{1,2}=0$ the charge stability diagram is given by the straight dashed lines in Fig.~\ref{fig:mips}(b). 
	However, it was observed experimentally~\cite{ferguson2023measurement} that the central straight line, separating 
	$\{01\}$ and $\{10\}$ 
	changes into an S-shape [green dot-dashed line in Fig.~\ref{fig:mips}(b)], which can be understood as a MIPS. 
	Population switching~\cite{PhysRevLett.79.127,baltin1999correlations,silvestrov2000towards} describes the occupation of a higher energy level at the expense of that of a lower energy level, due to the larger level width of the former.
	In the present experiment, the energy level more strongly coupled to the charge detector acquires a voltage-induced width $\Gamma_\varphi \propto \gamma V$, as in $A_{{\rm{WBL}}}$ above~\cite{ferguson2023measurement}. However, details of the experiment, encoded for example in a clearly observed asymmetry of the S-shape, $\delta S_1/\delta S_2$ 
	could not be reproduced by the phenomenological theory. We now show that our theory 
	not only 
	microscopically accounts for the MBA effects in this experiment [see Fig.~\ref{fig:mips}(b)], but also predicts some unexpected new features.

	As in Ref.~\onlinecite{ferguson2023measurement} we use rate equations $\partial_t P_i=\sum_j P_j\Gamma_{ji}-P_i\sum_j \Gamma_{ij}$  for $P_{i}$, $(i\in\{00,01,10,11\})$, the probabilities of the different charge states (summed over spin), in terms of the transition rates 
	$\Gamma_{ji}=\Gamma_{j \to i}$, see Fig.~\ref{fig:mips}(c), and solve for the steady state $\partial_t P_i=0$.  
	In Fig.~\ref{fig:mips}(b) we plot the contour  $P_{10}=P_{01}$, that marks the border between these two states. Consider for simplicity the case $\lambda_2=0$. We now apply Eq.~(\ref{eq:tun_rates}) for the tunneling rates (see Appendix \ref{sec:app_a} for a derivation),
	\begin{eqnarray}
		\label{eq:dd_tun_in}
		\Gamma^{({\rm{in}})}_{0n_2 \to 1 n_2}&=&s\Gamma \int dE\, f(E +\epsilon_1+U n_2 )A
		^+(E),\\
		\Gamma^{({\rm{out}})}_{1n_2 \to 0n_2}&=&\Gamma \int dE\,(1-f(E+\epsilon_1+U n_2))A
		^-(E),\nonumber 
	\end{eqnarray}
	where the factor of $s$ for tunneling-in rates is 
	$s=2$ $(s=1)$ for the spinful (spinless) case. For the tunnel events involving dot 2 the tunnel rates do not have a MBA effect if $\lambda_2=0$ (
	$\lambda_2>0$ can be treated similarly - see Appendix \ref{sec:app_A_3}). Our calculations in Fig.~\ref{fig:mips}(b)  give a good match of the experimental asmmetry~\cite{ferguson2023measurement}, $\delta S_1/\delta S_2=0.72 \pm 0.03$. 
	Below we describe more unique signatures beyond $P(E)$ theory.
	
	Along the diagonal dashed segment in Fig.~\ref{fig:mips}(b), for $\lambda_1\to 0$, the charge ground states are $\{10\}$ and $\{01\}$, and the lowest excited charge state changes from $\{00\}$ for $0> \epsilon_{1,2}> -U/2$ to $\{11\}$ for $-U/2 > \epsilon_{1,2} > -U$. The S-shape results from a competition of two 
	\textit{small}  rates~\cite{ferguson2023measurement}, 
	\bea
	\label{eq:smallrates}
	{\rm{For~}}0>\epsilon_{1,2}> -U/2:&&
	\Gamma^{({\rm{out}})}_{10\to 00} ~ {\rm{vs}} ~\Gamma^{({\rm{out}})}_{01\to 00}, \nonumber \\
	{\rm{For~}}-U/2 > \epsilon_{1,2} > -U:&&\Gamma^{({\rm{in}})}_{01\to 11} ~{\rm{vs}} ~\Gamma^{({\rm{in}})}_{10\to 11}.
	\eea
	In the region $0>\epsilon_{1,2}> -U/2$, since $\lambda_1>\lambda_2$, the tunneling-out rate is larger for $\{10\} \to \{00\}$, giving a preference for $\{01\}$; Likewise, in the region $-U/2 > \epsilon_{1,2} > -U$ the tunneling in rate dominates for $\{01\} \to \{11\}$, giving a preference to $\{10\}$. 

	
	First consider the spinless case effectively obtained at a large Zeeman field on the DD. 
	Indeed, for $s=1$ 
	the above two tunnel rates become identical at $\epsilon_1=-U/2$. For example, one can verify that $\Gamma_{01 \to 11}^{(\rm{in})}(\epsilon_1) = \Gamma_{10 \to 00}^{(\rm{out})}(-U-\epsilon_1)$,
	where $\epsilon \to -U-\epsilon$ is a reflection around the mid-valley point $\epsilon=-U/2$. Under this reflection, the charge states $\{10\}$ and $\{01\}$ exchange roles, together with the closest excited state. As a result, the S-shape would be exactly symmetric, passing through the point $\epsilon_1=\epsilon_2=-U/2$, in agreement with Ref.~\onlinecite{ferguson2023measurement}. 
	Thus, in the spinless case, any 
	difference 
	$\delta S_1 \ne \delta S_2$ is a smoking gun signature of MBA described by different functions $A^+(\omega) \ne A^-(-\omega)$, which goes beyond $P(E)$ physics. 
	Fig.~\ref{fig:mips}(d) depicts the ratio $S_1/\delta S_2$ vs $\epsilon_{\rm{d}}$ 
	for three values of the interaction $\lambda_1$ (here $\lambda_2$=0), demonstrating clear deviation from the $P(E)$ theory (straight broken line).  Note that in our setup, under $\epsilon_{\rm{d}} \to -\epsilon_{\rm{d}}$, $A^\pm(\omega)|_{\epsilon_{\rm{d}}} \to A^\mp(-\omega)|_{-\epsilon_{\rm{d}}}$ (see Appendix \ref{sec:app_B_1}) and consequently $\delta S_1$ and $\delta S_2$ gets interchanged.
	
	In the spinful case, as long as $A^+(E)=A^-(-E)$, due to the factor $s=2$ in Eq.~(\ref{eq:dd_tun_in}), $\{10\}$ is more prominent than $\{01\}$ at $\epsilon_{1,2}=-U/2$, giving $\delta S_1/\delta S_2 <1$.
	As a nontrivial prediction of AOC beyond the $P(E)$ theory, as seen in Fig.~\ref{fig:mips}(e), for strong enough $\lambda_1$, by changing $\epsilon_{d}$ one can change $\delta S_1/\delta S_2$ from below to above unity. \\

	\section{Outlook} 
	Our theory opens multiple future applications, both in the perturbative and non-perturbative regime in $\Gamma$. One important question that we leave for future study, is the regime of validity of the Maxwell relation route in entropy measurements~\cite{kuntsevich2015strongly,hartman2018direct,child2022entropy,piquard2023observing} which have promising applications~\cite{cooper2009observable,yang2009thermopower,viola2012thermoelectric,ben2013detecting,ben2015detecting,smirnov2015majorana,sela2019detecting,hou2012ettingshausen,han2022fractional,sankar2023measuring}. Secondly, our theory gives a starting point to explore nonequilibrtium effects on MBA-induced phase transitions~\cite{szyniszewski2019entanglement,kells2023topological,ma2023identifying}.

	\acknowledgements We gratefully acknowledge support from the European Research Council (ERC) under the European Union Horizon 2020 research and innovation programme under grant agreement
	No. 951541. ARO (W911NF-20-1-0013). Y.M. acknowledges support from ISF Grant
	No. 359/20. 
	The Flatiron Institute is a division of the Simons Foundation.
	
	\appendix

	\section{ Fermi's golden rule rates and spectral functions $A^{\pm}(\omega)$ \label{sec:app_a}}
	In this section we detail the derivation of the various detector assisted tunneling rates, Eqs.~(1), (6) and a generalization of Eq.~(8) , for different choices of the ``rest of the system" in Fig.~1(a) of the main text. \\
	
	\subsection{QD-lead system}
	Here we derive Eq.~(1) of the main text. Consider the spinless QD-lead system coupled to a charge detector. The detector is governed by some Hamiltonian, $H_n$, that depends on the charge $n=\{0,1\}$ of the system QD with energy level $\epsilon$. Let $|\psi_i^n\rangle$ denote the eigenbasis of $H_n$. The detector state is assumed to be given by the density matrix, $\rho_n=e^{-\beta H_n}/{\rm{Tr}} e^{-\beta H_n}=\sum_i P_i^{n}|\psi_i^n\rangle \langle \psi_i^n|$ where $P_i^{n}=\frac{e^{-\beta E_i^n}}{\sum_i e^{-\beta E_i^n}}$ and $\beta=T^{-1}$ is the inverse temperature. The tunnel rates given by the Fermi golden rule can be expressed as 
	\begin{widetext}
		\begin{eqnarray}
			\label{eq:Gammain}
			\Gamma^{{(\rm{in})}}&=&
			\Gamma\int dE f(E)\left[ \sum_{i,j}   
			P_j^0 \langle \psi_i^1| \psi_j^0\rangle \langle \psi_j^0|\psi_i^1 \rangle 2\pi \delta(E_j^0-E_i^1+E-\epsilon)\right],\\
			\Gamma^{{(\rm{out})}}&=&
			\Gamma\int dE (1-f(E))\left[ \sum_{i,j}P_i^1 
			\langle \psi_j^0| \psi_i^1\rangle \langle \psi_i^1|\psi_j^0 \rangle 2\pi \delta(E_i^1-E_j^0+\epsilon-E)\right].
		\end{eqnarray}
		Using $2\pi \delta(E)=\int\, dt e^{itE}$
		these expressions become
		\begin{eqnarray}
			\Gamma^{{(\rm{in})}}&=&\Gamma\int dE f(E)\int dt \, \left[ \sum_{i,j}
			P_j^0 \langle\psi_j^0|e^{itH_0}e^{-itH_1} |\psi_i^1 \rangle \langle \psi_i^1|\psi_j^0\rangle \right] e^{it(E-\epsilon)},\\
			\Gamma^{{(\rm{out})}}&=&\Gamma\int dE (1-f(E))\int dt \, \left[ \sum_{i,j}P_i^1 
			\langle \psi_i^1|e^{-itH_1}e^{itH_0} |\psi_j^0\rangle \langle\psi_j^0|\psi_i^1\rangle \right] e^{it(E-\epsilon)}.
		\end{eqnarray}
		Next we use $\sum_i 
		|\psi_i^1 
		\rangle \langle\psi_i^1|=\sum_j 
		|\psi_j^0 \rangle \langle\psi_j^0|=1$ to obtain
		\begin{eqnarray}
			\label{eq:apm}
			\Gamma^{{(\rm{in})}} &=&\Gamma\int dE f(E)\int dt \,\text{Tr}\left[\rho_0e^{iH_0t}e^{-iH_1t}\right]e^{it(E-\epsilon)}=\Gamma\int dE\, f(E)A^+(E-\epsilon),\\
			\Gamma^{{(\rm{out})}}&=&\Gamma\int dE(1- f(E))\int dt \,\text{Tr}\left[\rho_1e^{-iH_1t}e^{iH_0t}\right]e^{it(E-\epsilon)}=\Gamma\int dE\, (1-f(E))A^-(E-\epsilon).
		\end{eqnarray}
		We used the Lehman representation of the detector spectral functions, for example
		\bea
		A^+(E)&=& \int dt e^{i E t} {\rm{Tr}} \rho_0 e^{i H_0 t}e^{-i H_1 t}  \nonumber \\
		&=& \int dt \sum_{i,j} P^0_j  |\langle \psi^0_j |\psi^1_i \rangle|^2 e^{i E_j^0 t} e^{-i E_i^1 t}   e^{i E} = 2\pi \sum_{i,j} P^0_j |\langle \psi^0_j |\psi^1_i \rangle|^2 \delta(E_j^0-E_i^1+E).
		\eea
	\end{widetext}
	Thus, the rates of tunneling into the QD, $n =0 \to 1$, are accompanied by excitations in the detector as described by the spectral function $A^+$, and similarly for the tunneling out of the QD in terms of $A^-$. Note that the definitions of $A^\pm(E)$ in Eq.~(\ref{eq:apm}) are consistent with those in the main text
	\begin{equation}
		A^{\text{\tiny +/}\text{\small -}}(t)={\rm{Tr}}\left[\rho_{0/1}e^{itH_0}e^{-itH_1}\right],
	\end{equation}
	since the exponentials $^{itH_0},e^{-itH_1}$ do commute inside this trace (one of them commutes with $\rho_{0/1}$).  Note also that it follows that $A^{\pm}(-t)=\left(A^{\pm}(t)\right)^*$. 
	

	\subsection{Tunnel coupled double dot}
	Here we derive Eq.~(6) of the main text. Consider a double dot Hamiltonian $H_{DD}=\epsilon_1 c^\dagger_1 c_1+\epsilon_2 c^\dagger_2 c_2+ (v c^\dagger_1 c_2 + h.c.)$ where $n=d^\dagger_1 d_1$. Using the exact same treatment we obtain for the tunneling rate into dot 1
	\begin{equation}
		\Gamma^{{(\rm{in})}}_{{\rm{DD}}}=2\pi |v|^2  \sum_{i,j} 
		P_j^0 \langle \psi_i^1| \psi_j^0 \rangle \langle \psi_j^0|\psi_i^1 \rangle \delta(E_j^0-E_i^1+\epsilon_2-\epsilon_1),
	\end{equation}
	which is the same as Eq.~(\ref{eq:Gammain}) with $E \to \epsilon_2, \epsilon \to \epsilon_1$, with no Fermi function (since we consider the conditional transition probability to tunnel into dot 1 given that an electron is in dot 2), and no integral over $E$. As a result we obtain
	\be
	\Gamma^{({\rm{in}})}_{{\rm{DD}}}=|v|^2 A^+(\epsilon_2-\epsilon_1),~~~\Gamma^{{(\rm{out})}}_{{\rm{DD}}}=|v|^2 A^-(\epsilon_2-\epsilon_1).
	\ee
	Namely, by scanning the DD level detuning, a measurement of the tunneling rates directly measures the detector spectral functions.
	
	\subsection{Capacitively coupled double dot \label{sec:app_A_3}}
	Here we provide the full expression for the tunneling rates in the system in Fig.~3(a) of the main text for finite $\lambda_1$ and $\lambda_2$, generalizing Eq.~(8) in the main text.  For finite $\lambda_1$ and $\lambda_2$  the detector Hamiltonian takes one out of four possible values, $H_{n_1,n_2}=H_{{\rm{det}}}+H_{\lambda}(n_1,n_2)$ and we define for the tunneling into or out of dot 1
	\bea
	A^{+}_{1,n_2}={\rm{Tr}} [e^{-\beta H_{0,n_2}} e^{i t H_{0,n_2}}  e^{-i t H_{1,n_2}}],  \\
	A^{-}_{1,n_2}={\rm{Tr}} [e^{-\beta H_{1,n_2}} e^{i t H_{0,n_2}}  e^{-i t H_{1,n_2}}], \nonumber
	\eea
	and similarly for tunneling in or out of dot 2,
	\bea
	A^{+}_{2,n_1}={\rm{Tr}} [e^{-\beta H_{n_1,0}} e^{i t H_{n_1,0}}  e^{-i t H_{n_1,1}}],  \\
	A^{-}_{2,n_1}={\rm{Tr}} [e^{-\beta H_{n_1,1}} e^{i t H_{n_1,0}}  e^{-i t H_{n_1,1}}]. \nonumber
	\eea
	Now the tunneling in or out of dot 1 are given by
	\begin{equation}
		\label{eq:dd_tun_in}
		\Gamma^{({\rm{in}})}_{0n_2 \to 1 n_2}=s\Gamma \int d\omega\, f(\omega)A_{1,n_2}^+(\omega-\epsilon_1-U n_2),
	\end{equation}
	\begin{equation}
		\Gamma^{({\rm{out}})}_{1n_2 \to 0n_2}=\Gamma \int d\omega\,f(-\omega)A_{1,n_2}^-(\omega-\epsilon_1-U n_2),
	\end{equation}
	where the factor in tunneling-in rates is $s=1(2)$ for the spinless (spinful) case. Similarly, 
	\begin{equation}
		\Gamma^{({\rm{in}})}_{n_10 \to  n_1 1}=s\Gamma \int d\omega\, f(\omega)A_{2,n_1}^+(\omega-\epsilon_2-U n_1),
	\end{equation}
	\begin{equation}
		\Gamma^{({\rm{out}})}_{n_1 1 \to n_1 0}=\Gamma \int d\omega\,f(-\omega)A_{2,n_1}^-(\omega-\epsilon_2-U n_1).
	\end{equation}

	\section{ Exact numerical method \label{sec:app_B}}
	In this section we provide details for our numerically exact method to compute the correlators $A^{\pm}(t)$ 
	when the detector Hamiltonians, $H_n$, are non-interacting.
	We focus on the QDD model, 
	\begin{eqnarray}
		H_n&=&H_{{\rm{det}}} + \lambda \left(d^\dagger d -\frac{1}{2}\right)\left(n-\frac{1}{2}\right),\\
		H_{\rm{det}}&=&\epsilon_{\rm{d}} d^\dagger d  + \sum_{k,\mu=L,R}\epsilon_k \varphi_{k \mu}^\dagger \varphi_{k \mu} \nonumber \\
		& &+ \sum_{\mu=L/R} v_\mu \sum_k(d^\dagger \varphi_{k,\mu} + h.c. ).
	\end{eqnarray}
	The QDD leads are voltage biased with $\pm V/2$. 
	We assume that the QDD leads have a constant DoS, $\nu$, and then the line width function is $\Gamma_{d} = 2\pi \nu(v_L^2+v_R^2) $. 
	The local DoS of the QDD dot is then given by the Lorentzian form,
	\begin{equation}
		\nu_{d}(\omega;\epsilon_{\rm{d}}) = \frac{1}{\pi}\frac{\Gamma_{d}}{(\omega-\epsilon_{\rm{d}})^2+\Gamma_{d}^2}.
	\end{equation}
	The lesser and greater Green functions of the QDD with respect to $H_{\rm{det}}$ are given by
	\begin{equation}
		\label{eq:qdd_great}
		g^>(\omega;\epsilon_{\rm{d}})=2\pi i \,\nu(\omega;\epsilon_{\rm{d}})[f(\omega-V/2)+f(\omega+V/2)-2],
	\end{equation}
	\begin{equation}
		\label{eq:qdd_less}
		g^<(\omega;\epsilon_{\rm{d}})=2\pi i\, \nu(\omega;\epsilon_{\rm{d}})[f(\omega-V/2)+f(\omega+V/2)].
	\end{equation}
	Using the linked-cluster theorem we can write $  A^{\pm}(t)=e^{C^{\pm}(t)}$, where $C^{\pm}(t)$ is the sum of all single loop diagrams \cite{nozieres1969singularities},
	\begin{widetext}
		\begin{equation}
			\label{eq:cpm}
			C^\pm(t)=-\sum_{k= 1}^\infty\frac{(\pm \lambda)^k}{k}\int_0^{\pm t} dt_1\ldots \int_0^{\pm t} dt_k \, g^{(n)}(t_1-t_2)\ldots g^{(n)}(t_k-t_1),
		\end{equation}
	\end{widetext}
	with 
	$n=\{0,1\} $  for $C^\pm$, respectively, and $g^{(n)}(t_1-t_2)$ denotes the time-ordered Green function with respect to $H_n$,
	\begin{equation}
		g^{(n)}(t_1-t_2)=-i \text{Tr}\left[\rho_n \mathcal{T} d(t_1) d^\dagger(t_2) \right].
	\end{equation}
	
	We now describe the calculation of $C^+(t)$ (with a similar calculation of $C^-(t)$), and for simplicity we  drop the 
	superscripts of $g^{(n)}$ such that $g \equiv g^{(0)} $. To proceed, it is useful to define  a special Green function $\psi_t(u,u')$, which is the sum of all connected diagrams with two external legs at times $u$ and $u'$ and the vertices are restricted to lie in the range $[0,t]$. It obeys a quasi-Dyson equation,
	\begin{equation}
		\label{eq:psi}
		\psi_t(u,u')=g(u-u')+\lambda \int_0^t dt_1\, g(u-t_1)\psi_t(t_1,u').
	\end{equation}
	Comparing Eqs.~\eqref{eq:cpm} and \eqref{eq:psi} we get
	\begin{equation}
		\label{eq:dif_c}
		\frac{dC^+}{dt}=-\lambda \psi_t(t,t),~~~C^+(0)=0.
	\end{equation}

	The numerical method proceeds in two steps: the first step is to obtain $\psi_t(u,u')$ by solving the quasi-Dyson equation and the second step is to integrate Eq.~\eqref{eq:dif_c}. Since we only need the special case, $u'=t$, it suffices to solve for $\psi_t(u,t)$, which can be expressed by splitting the time ordered Green functions into lesser/greater Green functions as
	\begin{align}
		\psi_t(u,t)&=&g^<(u-t)+\lambda \left[\int_0^{u} dt_1\, g^>(u-t_1)\psi_t(t_1,t) \right. \nonumber\\
		& &\left.+\int_u^t dt_1\,g^<(u-t_1)\psi_t(t_1,t)\right].
	\end{align}
	The functions $g^>(t)$ and $g^<(t)$ are respectively obtained by taking a numerical inverse Fourier-transform of Eq.~\eqref{eq:qdd_great} and \eqref{eq:qdd_less} with $\epsilon_{\rm{d}}$ replaced by $\epsilon_{\rm{d}} - \lambda / 2$.
	Discretizing the time-interval, $[0,t]$, to a regular grid of step $dt=t/N$: $\{u_0,\hdots,u_N\}$, and denoting $\psi_t(u_i,t) \to h(u_i)$, the above equation becomes
	\begin{align}
		\label{eq:hui}
		h(u_i)&=&g^<(u_i-t)+\lambda \left[ \sum_{r=0}^{i-1} \int_{u_r}^{u_{r+1}}dv\, g^>(u_i-v)h(v) \right. \nonumber \\
		& & \left.+\sum_{r=i}^{N-1}\int_{u_r}^{u_{r+1}}dv\, g^<(u_i-v)h(v)\right].
	\end{align}
	The integrals can be performed by any quadrature and the discretization of the grid is chosen according to the desired accuracy. Specifically we choose the following quadrature, written compactly as,
	\begin{align}
		\int_{u_r}^{u_{r+1}}dv\, g(u_i-v)h(v)=\frac{dt}{3}(g_{i-r-1}h_{r+1}+g_{i-r}h_r)\nonumber \\
		+\frac{dt}{6}(g_{i-r-1}h_r+g_{i-r}h_{r+1}) + O(dt^2),
	\end{align}
	which is the result of approximating $g$ and $h$ locally as linear functions.
	Eq.~\eqref{eq:hui} then becomes a linear problem $\sum_j M_{ij} h(u_j) = g^<(u_i - t)$ which is solved efficiently with the GMRES algorithm.
	We exploit the fact that the matrix $M$ is almost of Toeplitz form (up to two rows and two columns) to perform fast matrix-vector product with the Fast Fourier Transform algorithm.
	
	Note that  to obtain $\psi_t(u,t)$ for different $t$, a linear problem has to be solved separately for each $t$. For this, another grid discretization is applied in the interval $[0,t_{max}]$, with $t_{max}$ chosen such that $A^{\pm}(t>t_{max})$ is given by a simple exponential decay, which always happens due to a finite bias or temperature. 
	The step size is defined adaptively to ensure that a desired accuracy is reached all along the interval.
	The function $\psi_{t}(t,t)$ obtained by solving the quasi-Dyson equation for each $t$ in the grid is then used to numerically integrate Eq.~\eqref{eq:dif_c} with a Simpson rule. Finally a numerical Fourier transform is applied for $A^{\pm}(t)$ to obtain $A^{\pm}(\omega)$, taking care to properly account for the exponential decay for $t>t_{max}$. 
	
	We typically neglect any Hartree shift. It enters as Im$(C^\pm(t))=\Delta_{H} t$ in the calculation. We simply subtract out this linear contribution and then  take the numerical Fourier transform. 
	
	\subsection{ Particle-hole transformation \label{sec:app_B_1}} 
	Here we show that, a particle-hole transformation relates $A^{\pm}(t)$ for  QDD levels  $\epsilon_{\rm{d}}$ and $-\epsilon_{\rm{d}}$.
	It is convenient to express $A^+(t)$ using the many-body formulation
	\bea
	A^+(t)= {\rm{Tr}} \{ \rho_0 e^{i H_0 t} e^{-i H_1 t} \}={\rm{Tr}}\{ e^{-\beta H} c(t) c^\dagger \},
	\eea
	where $H=H_0 c c^\dagger+H_1 c^\dagger c$.
	Now we want to express 
	\bea
	A^+(t)|_{-\epsilon_{{\rm{d}}}}={\rm{Tr}}\{ e^{-\beta H |_{-\epsilon_{{\rm{d}}}}} c(t) c^\dagger \}
	\eea
	in terms of the detector spectral functions at $+\epsilon_{{\rm{d}}}$. For this purpose, we apply the particle hole transformation
	\be
	d \to d^\dagger,~~~c \to c^\dagger,~~~\epsilon_{{\rm{d}}} \to - \epsilon_{{\rm{d}}}.
	\ee
	This gives
	\bea
	A^+(t)|_{-\epsilon_{{\rm{d}}}}&=&{\rm{Tr}}\{ e^{-\beta H |_{+\epsilon_{{\rm{d}}}}  } c^\dagger(t) c \} \nonumber \\
	&=&{\rm{Tr}}\{ e^{-\beta H |_{+\epsilon_{{\rm{d}}}}  } c^\dagger(0) c(-t) \} \nonumber \\ 
	&=& A^-(-t)|_{\epsilon_{{\rm{d}}}}.
	\eea
	In the last equation we used the many-body formulation of $A^-(t)$,
	\bea
	A^-(t)= {\rm{Tr}} \{ \rho_1  e^{i H_0 t} e^{-i H_1 t}   \}={\rm{Tr}}\{ e^{-\beta H} c^\dagger(0) c(t) \}.
	\eea
	Therefore
	\begin{align}
		A^+(E)|_{-\epsilon_{{\rm{d}}}} =\int dt e^{i E t}A^+(t)|_{-\epsilon_{{\rm{d}}}}=\int dt e^{i E t}A^-(-t)|_{+\epsilon_{{\rm{d}}}}  \nonumber \\
		= \int dt e^{-i E t}A^-(t)|_{+\epsilon_{{\rm{d}}}}=A^-(-E)|_{+\epsilon_{{\rm{d}}}}.
	\end{align}

	\section{Analytical form in the wide band limit \label{sec:aap_C}}
	In this section, we refer to analytic formulas for $A^\pm(t)$ in the wide-band-limit (to keep notation concise, we drop the subscript $\rm{WBL}$ in all subsequent expressions)~\cite{aleiner1997dephasing}
	\begin{equation}
		\label{eq:a_time_v}
		A^{\pm}(t)=\left(\frac{\pi T}{\pm i \xi \sinh(\pi T t) }\right)^{\alpha}e^{-\Gamma_\varphi |t|+\gamma h(t)},~~~( t \gg 1/\xi). 
	\end{equation}
	Here the high energy cut-off $\xi \sim \Gamma_{\rm{d}}$ is introduced to cure the short-time divergence. 
	The function $h(t)$ is given by $h(t)=\int_{0}^{t}d\tau\frac{\pi^2T^2 \tau (1-\cos(V\tau))}{\sinh^2(\pi T \tau)}$. We can see that in the WBL $A^-(t)= A^+(-t)$ (and $A^\pm (-t)=(A^\pm (t))^*$   by definition) yielding $A^+(\omega)=A^-(\omega)$. As emphasized, this is not a general property, which is violated by finite bandwidth effects such as finite $\Gamma_{\rm{d}}$. 
	Taking the Fourier transform neglecting $h(t)$, we get,
	\begin{equation}
		\label{eq:agreat_w_v_anlyt}
		A^{\pm}(\omega) \xrightarrow{|\omega|<\xi} \chi(\alpha) \cdot \, {\rm{Re}}\left[\frac{e^{\pm i\pi\alpha/2}\Gamma(\frac{i\omega+\Gamma_\varphi}{2\pi T}+\frac{\alpha}{2})}{\Gamma(1+\frac{i\omega+\Gamma_\varphi}{2\pi T}-\frac{\alpha}{2})}\right], 
	\end{equation}
	with $\chi(\alpha)=\frac{2^\alpha}{\xi}\left(\frac{\pi T}{\xi}\right)^{\alpha-1}\Gamma(1-\alpha)$.

	\section{ Perturbative calculation of $A^\pm(t)$ and connection to $P(E)$ theory \label{sec:app_D} }
	We consider a generic model~\cite{aleiner1997dephasing} where the detector is governed by a Hamiltonians $H_n$ depending on the charge state $n=\{0,1\}$ of the QD, 
	\bea
	H_0&=&\sum_k \epsilon_k \left[L_k^\dagger L_k +R_k^\dagger R_k\right], \nonumber \\
	H_1&=&H_0+\lambda_{1} \sum_{k_1,k_2}(L_{k_1}^\dagger L_{k_2}+R^\dagger_{k_1}R_{k_2}) \nonumber \\
	& &+ \lambda_{LR}\sum_{k_1,k_2} (L^\dagger_{k_1}R_{k_2}+ h.c.),
	\eea
	written in the basis of scattering eigenstates for the case, $n=0$. Here, for simplicity, we consider the spinless case and so $m=1$. The $L$ and $R$ states have chemical potentials $\mu_L,\mu_R$ with $\mu_R-\mu_L=V$. For the $n=1$ charge state, the additional potential at the detector induces both forward scattering and backscatternig between the eigenstates of $H_0$. Eq.~(\ref{eq:a_time_v}) was derived within this model. 
	Below we outline the essence of that calculation emphasizing the relation to $P(E)$ theory.
	
	We have $A^+(t)=e^{C^+(t)}$, with $C^+(t)$ given by Eq.~\eqref{eq:cpm}. We are interested in calculating $C^+(t)$ up to quadratic order in the scattering coefficients. Note that the first order term could lead to a Hartree shift, which we ignore. The second order term is (we drop the $+$ superscript and use the shortcut, $\int_1=\int_0^tdt_1$),
	\begin{widetext}
		
		\begin{equation}
			C^{(2)}(t)=-\frac{\lambda_1^2}{2}\sum_{i=L/R}\int_1\int_2 g_{i}(t_1-t_2)g_{i}(t_2-t_1)-\frac{\lambda_{LR}^2}{2}\left[\int_1 \int_2 g_L(t_1-t_2)g_R(t_2-t_1)+\int_1 \int_2 g_R(t_1-t_2)g_L(t_2-t_1)\right].
		\end{equation}
	\end{widetext}
	We now write the time-ordered Green functions in terms of greater and lesser Green functions,
	\begin{equation}
		g(\tau)=\theta(\tau)g^>(\tau)+\theta(-\tau)g^<(\tau),
	\end{equation}
	and obtain,
	\begin{equation}
		C^{(2)}(t)=-\sum_{\{i,j\}=\{L,R\}} \lambda_{ij}^2 \int_0^t dt_1\int_0^{t_1}dt_2\,g_i^>(t_1-t_2)g_j^<(t_2-t_1),
	\end{equation}
	where we defined $\lambda_{LL}=\lambda_{RR}=\lambda_1$ and $\lambda_{RL}=\lambda_{LR}$. Next, writing the time domain Green functions as inverse Fourier transforms of the frequency domain ones, we obtain,
	
	\begin{align}
		\label{eq:c2_inter_1}
		C^{(2)}(t)=-\sum_{\{i,j\}=\{L,R\}} \frac{\lambda_{ij}^2}{4\pi^2}\left[\int_{-\infty}^{\infty} d\Omega\, \chi_{ij}(\Omega) \right. \nonumber \\
		\left.\int_0^t dt_1 \int_0^{t_1}dt_2\, e^{i\Omega(t_1-t_2)}\right],
	\end{align}
	where
	\begin{equation}
		\label{eq:chi_def}
		\chi_{ij}(\Omega)=\int_{-\infty}^{\infty} d\omega\, g_i^>(\omega)g_j^<(\omega+\Omega).
	\end{equation}
	The time integrals in Eq.~(\ref{eq:c2_inter_1}) give
	\begin{equation}
		\int_0^t dt_1 \int_0^{t_1}dt_2\, e^{i\Omega(t_1-t_2)}=\frac{it}{\Omega}+\frac{1-e^{i\Omega t}}{\Omega^2}.
	\end{equation}
	The first term only contributes to the Hartree shift and we neglect it. Then,
	\begin{equation}
		\label{eq:c2t_main_form}
		C^{(2)}(t)=-\sum_{\{i,j\}=\{L,R\}} \frac{\lambda_{ij}^2}{4\pi^2}\int_{-\infty}^{\infty} d\Omega\, \chi_{ij}(\Omega) \frac{1-e^{i\Omega t}}{\Omega^2}.
	\end{equation}
	In the wide band limit  with an energy independent DoS, $\nu$, the local Green function is
	\begin{equation}
		g_j^>(\omega)=-2\pi \nu i(1-f(\omega-\mu_j)),\quad g_j^<(\omega)=2\pi \nu i f(\omega-\mu_j).
	\end{equation}
	Plugging it in Eq.~(\ref{eq:chi_def}) we obtain
	\begin{equation}
		\label{eq:chi_form}
		\chi_{ij}(\Omega)=4\pi^2\nu^2 \frac{\Omega+\mu_i-\mu_j}{-1+\exp[\frac{\Omega+\mu_i-\mu_j}{T}]}.
	\end{equation}
	We now make the connection to the $P(E)$ description in Ref.~\onlinecite{aguado2000double}. For this, compare our expressions in Eqs.~\eqref{eq:c2t_main_form} and~\eqref{eq:chi_form} with analogous expressions, equations (3) and (7) in Ref.~\onlinecite{aguado2000double}. They are identical once the impedance function $Z(\omega)$ is assumed to be frequency independent (which is usually assumed in the applications of $P(E)$ in literature) and the scattering matrix elements are appropriately redefined. Thus we conclude that the $P(E)$ theory, as considered in Ref.~\onlinecite{aguado2000double}, is the perturbative limit of our more general theory where the AOC is treated exactly.

	Now we proceed with the evaluation of the integral in Eq.~\eqref{eq:c2t_main_form}. Let us consider the case of $T=0$, for which Eq.~\eqref{eq:chi_form} reduces to
	\begin{equation}
		\chi_{ij}(\Omega) \xrightarrow{T=0} 4\pi^2\nu^2(\mu_j-\mu_i-\Omega) \Theta(\mu_j-\mu_i-\Omega).
	\end{equation}
	Let us split, $C^{(2)}(t)=C_1(t)+C_2(t)$ where $C_1(t)=C_{LL}(t)+C_{RR}(t)$ and $C_2(t)=C_{LR}(t)+C_{RL}(t)$. Let us also define $g_{1/LR}=\nu \lambda_{1/LR}$. We get,
	\begin{equation}
		C_1(t)=2g_1^2 \int_{-\xi}^0 d\Omega\,\frac{1-e^{i\Omega t}}{\Omega} \xrightarrow{t \gg 1/\xi} -2g_1^2\log(\xi t),
	\end{equation}
	where the high-energy cutoff, $\xi$, was introduced to cure the ultra-violet divergence. This term leads to the equilibrium AOC exponent,
	\begin{equation}
		C^{(2)}(t) = -\alpha \log(t),
	\end{equation}
	with,
	\begin{equation}
		\alpha=2g_{1}^2.
	\end{equation}
	
	For the second term, we have, 
	\begin{widetext}
		\begin{eqnarray}
			C_2(t) &=& -g_{LR}^2\left[\int_{-\infty}^V d\Omega\,\frac{V-\Omega}{\Omega^2}(1-e^{i\Omega t})+ \int_{-\infty}^{-V}d\Omega\, \frac{-V-\Omega}{\Omega^2}(1-e^{i\Omega t})\right]\\
			&=& -g_{LR}^2\left[V\int_{-V}^{V}d\Omega\,\frac{1-e^{i\Omega t}}{\Omega^2}-2\int_{-\infty}^{-V}d\Omega\,\frac{1-e^{i\Omega t}}{\Omega}-\int_{-V}^{V}d\Omega\,\frac{1-e^{i\Omega t}}{\Omega}\right]\\
			&=&-g_{LR}^2\left[2V\int_0^V d\Omega\,\frac{1-\cos\Omega t}{\Omega^2}-2\int_{-\xi}^0 d\Omega\,\frac{1-e^{i\Omega t}}{\Omega}+2\int_{-V}^0 d\Omega\,\frac{1-e^{i\Omega t}}{\Omega}+2i\int_0^V d\Omega\, \frac{\sin \Omega t}{\Omega} \right].
		\end{eqnarray}
	\end{widetext}
	The imaginary part of the third last term cancels the fourth term and we then have,
	\begin{equation}
		C_2(t)=-2g_{LR}^2 \left[A+B+C\right].
	\end{equation}
	Let us evaluate each term separately:
	\begin{eqnarray}
		A&=&Vt\int_0^{Vt} dx\,\frac{1-\cos x}{x^2}  \xrightarrow{t\gg 1/V} \frac{\pi}{2}Vt,\\
		B&=&-\int_{-\xi t}^0 dx\, \frac{1-e^{ix}}{x} \xrightarrow{t\gg 1/\xi} \log(\xi t),\\
		C&=&-\int_{0}^{Vt} dx\, \frac{1-\cos x}{x} .
	\end{eqnarray}
	Note that the $C$ integral corresponds to the function $h(t)$ of Eq.~(\ref{eq:a_time_v}). Crucially, from this calculation we identify the linear term corresponding to the dephasing rate
	\begin{equation}
		C_2(t)\xrightarrow{t\gg 1/V} -\pi g_{LR}^2Vt = -\Gamma_\varphi t,
	\end{equation}
	with 
	\begin{equation}
		\Gamma_\varphi=\pi g_{LR}^2 V.
	\end{equation}

	\section{Parameters used in the numerical calculation of Fig. 3b in main text \label{sec:app_E}}
	The parameters of the capacitively coupled double dot of Ref.~\onlinecite{ferguson2023measurement}, in addition to the temperature $T$ consist of: 
	\begin{itemize}
		\item  System parameters: $(\epsilon_1,\epsilon_2,U,\Gamma)$. The experiment is in the regime $\Gamma \ll T$, $U/T$ is known experimentally and the energy levels are scanned using a gate voltage in units of $U$.
		\item Interaction parameters $(\lambda_1,\lambda_2)$. In the experiment a Lorentzian broadening $\propto \lambda_i^2$ was assumed, and the asymmetry between the broadening due to $\lambda_1$ and $\lambda_2$ was estimated to be four, hence we set $\lambda_1/\lambda_2\approx 2$.
		\item Detector parameters ($\epsilon_{\rm{d}},\Gamma_{\rm{d}},V$).  We consider transmission function of the QDD, $\mathcal{T}=\Gamma_{\rm{d}}^2/(\epsilon_{\rm{d}}^2+\Gamma_{\rm{d}}^2)$ and demand that its change due to change in charge state of QD1 is around $\Delta \mathcal{T}/\mathcal{T}\approx 20\%$  as reported in the experiment (Fig. 9 in Ref.~\onlinecite{ferguson2023measurement}). We then select $\epsilon_{\rm{d}}=0.25\Gamma_{\rm{d}}$ and $\lambda_1=0.5 \Gamma_{\rm{d}}$, which leads to a similar change in transmission.
	\end{itemize}
	The typical voltage and temperature in the experiment satisfy 
	\begin{equation}
		V\sim U, \quad T\sim 0.025\,U.
	\end{equation}
	
	The value of $\Gamma_{\rm{d}}$ is not known from experiment, but we find that its value is not important. In the main text we considered $\Gamma_{\rm{d}}=U$. Our result is in  excellent agreement with the experimentally observed asymmetry, $\delta S_1/\delta S_2$ as discussed in the main text.

\end{document}